\documentclass[preprintnumbers,secnumarabic,showpacs,amsmath,amssymb]{revtex4}
\usepackage{docs}

\usepackage{graphicx}
\usepackage{dcolumn}

\usepackage{bm}
\begin{document}
\title{Van der Waals quintessence stars}%


\author{Francisco S. N. Lobo}%
\email{flobo@cosmo.fis.fc.ul.pt}
\affiliation{Centro de Astronomia
e Astrof\'{\i}sica da Universidade de Lisboa, Campo Grande, Ed. C8
1749-016 Lisboa, Portugal}

\affiliation{Institute of Gravitation \& Cosmology, University of
Portsmouth, Portsmouth PO1 2EG, UK}

\begin{abstract}

The van der Waals quintessence equation of state is an interesting
scenario for describing the late universe, and seems to provide a
solution to the puzzle of dark energy, without the presence of
exotic fluids or modifications of the Friedmann equations. In this
work, the construction of inhomogeneous compact spheres supported
by a van der Waals equation of state is explored. These
relativistic stellar configurations shall be denoted as {\it van
der Waals quintessence stars}. Despite of the fact that, in a
cosmological context, the van der Waals fluid is considered
homogeneous, inhomogeneities may arise through gravitational
instabilities. Thus, these solutions may possibly originate from
density fluctuations in the cosmological background. Two specific
classes of solutions, namely, gravastars and traversable wormholes
are analyzed. Exact solutions are found, and their respective
characteristics and physical properties are further explored.

\end{abstract}


\maketitle

\section{Introduction}

The universe is presently undergoing an accelerated phase of
expansion. Several candidates have been proposed in the literature
to explain this cosmic accelerated expansion, such as dark energy
models, the generalized Chaplygin gas, modified gravity and
scalar-tensor theories, tachyon scalar fields and specific
braneworld models, such as the Dvali-Gabadadze-Porrati (DGP)
model. However, it was pointed out that assuming that dark energy
is governed by a perfect fluid equation of state may
systematically induce wrong results and be misleading in inferring
the nature of dark energy \cite{Capo2}. In this spirit, an
alternative model was recently proposed without the presence of
exotic fluids and modifications of the Friedmann equations, using
a more complicated equation of state, namely, the van der Waals
(VDW) equation of state~\cite{Capoa, Capob}, given by
\begin{equation}
p=\frac{\gamma \rho}{1-\beta \rho}-\alpha \rho^2  \,,
\label{vdWstar1st}
\end{equation}
where $\rho$ is the energy density and $p$ the pressure of the VDW
fluid. The accelerated and decelerated periods depend on the
parameters, $\alpha$, $\beta$ and $\gamma$ of the equation of
state, and in the limiting case $\alpha,\beta \rightarrow 0$, one
recovers the dark energy equation of state, with $\gamma=p/\rho
<-1/3$. It was also stressed that the perfect fluid equation of
state $p=\gamma \rho$ reflects an approximation of cosmic epochs
describing stationary situations, where phase transitions are not
considered \cite{Capo3}. Thus, one of the advantages of the VDW
model is that it describes the transition from a matter field
dominated era to a scalar field dominated epoch, without
introducing scalar fields. Furthermore, it is useful to explain
the universe with a minimal number of ingredients, and the VDW gas
actually treats dark matter and dark energy as a single fluid. The
VDW quintessence scenario has also been confronted successfully
with a wide variety of observational tests, by constraining the
free parameters \cite{Capo3}.

The success of this model has stimulated several different
approaches. Recently, a model with a binary mixture whose
constituents are described by a VDW fluid and by a dark energy
density was also proposed \cite{Kremer}. It is interesting to note
that this model can simulate several aspects, namely, an
inflationary period where the acceleration grows exponentially;
and a present accelerated period where the dark energy density
dominates over the energy density of the VDW fluid. The
construction of a general scheme where the VDW ideal fluid has a
mathematically equivalent representation as a scalar-tensor theory
with a specific potential, as well as explicit examples were also
explored in Ref. \cite{CNO}. Another approach has been analyzed in
the context of brane cosmology \cite{Kremer2}, where the
cosmological fluid on the brane is modeled by the VDW equation of
state. It was shown that this model reproduces several features,
namely, an initial accelerated epoch where the VDW fluid behaves
like a scalar field with a negative pressure, and a present
accelerated phase due to a cosmological constant on the brane.

Despite of the fact that the VDW equation of state represents a
spatially homogeneous cosmic fluid and is assumed not to cluster,
due to gravitational instabilities inhomogeneities may arise.
Thus, the solutions outlined in this paper may possibly originate
from density fluctuations in the cosmological background,
resulting in the nucleation through the respective density
perturbations. Therefore, the pressure in the VDW equation of
state may be regarded a radial pressure, and the transverse
pressure is determined through the Einstein field equations.
The justification of this approach may be considered as an
extension of the construction of inhomogeneous wormhole spacetimes
supported by phantom energy~\cite{phantomWH}. The latter
construction was motivated by the analysis carried out in Ref.
\cite{Sush-Kim}, where time-dependent solutions were constructed
with ghost scalar fields. In this model, it was shown that the
radial pressure is negative everywhere and equals the transverse
pressure far from the throat, showing that ghost scalar fields
behave essentially as dark energy. Now, in the VDW context, one
may also consider, in principle, that for large radial distances
in these inhomogeneous solutions the radial pressure equals the
transverse pressure, and thus behaving essentially as a VDW
cosmological fluid, consequently justifying the above approach.
The transverse pressure obtained from the field equations should
not be considered typical of the cosmic fluid driving the
accelerated expansion, as it is only a feature of these
inhomogeneous VDW configurations.
It is also important to emphasize that the equation of state
leading to the acceleration of the Universe on large scales, is an
average equation of state corresponding to a background fluid. So
in the inhomogeneities giving rise to the configurations discussed
in this work, it is also important to justify if one may really
use an equation of state that in fact corresponds to a background
fluid. For this, it is important to understand the evolution of
structure formation in the VDW scheme. It was shown that for large
redshifts the VDW model is similar to the $\Lambda$CDM model
\cite{Capo3}, where the VDW dark matter plays the role of a
cosmological constant, showing that structure formation could
evolve in a similar manner as in the $\Lambda$CDM model, although
these points need to be further investigated~\cite{Capo3}. Now, it
was also shown that due to a negative adiabatic sound speed,
perturbations are unstable in the VDW quintessence scenario, and
inevitably the role of anisotropic stresses and entropy production
has to be taken into account. Thus, as a first approximation and
as a working hypothesis, one may argue that this also justifies
the above approach of using an equation of state that corresponds
to a background fluid, with a radial pressure.
We shall denote these inhomogeneous spherically symmetric
solutions as {\it van der Waals quintessence stars}. It is also
interesting to note that the thermodynamic properties of a general
relativistic isotropic pressure self-gravitating ball of fluid,
undergoing a generic first order phase transition and governed by
the VDW equation of state, was studied in Ref. \cite{Polanco}. In
this work, we shall generalize the latter to an anisotropic
pressure general relativistic model, and we shall consider two
specific class of solutions of VDW quintessence stars, namely,
gravastars and traversable wormholes.

The ``gravastar'' ({\it grav}itational {\it va}cuum {\it star})
picture developed by Mazur and Mottola \cite{Mazur}, initially
arouse from a considerable amount of scepticism related to
spacetime singularities and event horizons in the Schwarzschild
black hole solution (see Ref. \cite{Doran} for a review on
specific misconceptions and ambiguities related to these issues).
In the gravastar model, the interior Schwarzschild solution is
replaced with a de Sitter condensate, and thus, does away with the
problem of the singularity at the origin and the event horizon. In
this model, the quantum vacuum undergoes a phase transition at or
near the location where the event horizon is expected to form. In
this spirit, it is interesting to note that several models have
been proposed to some extent in the literature \cite{Dymnikova}.
Recently, motivated by dark energy quintessence, which is a
possible candidate responsible for the late-time cosmic
accelerated expansion, a generalization of the gravastar model has
been proposed \cite{darkstar1,darkstar2}, by considering an
interior solution governed by the dark energy equation of state,
$\gamma=p/\rho<-1/3$. Stable dark energy stellar models were
analyzed in Ref. \cite{darkstar2}, and it was found that large
stability regions exist that are sufficiently close to where the
event horizon is expected to form. It was also emphasized that it
would be difficult to distinguish the exterior geometry of the
dark energy stars from an astrophysical black hole. In this
context, it is interesting to note that stars supported by a
generalized Chaplygin gas equation of state have been explored
\cite{Chapstar}.

In this work, we shall also be interested in investigating the
construction of traversable wormholes \cite{Morris} supported by
the VDW equation of state. An analogous approach was recently
carried out in the presence of phantom energy \cite{phantomWH} and
the generalized Chaplygin gas \cite{Lobo-chaplygin}. We will show
that traversable wormhole solutions may be constructed using the
VDW equation of state, which are either asymptotically flat or
possess finite dimensions, where the exotic matter is confined to
the throat neighborhood. The latter solutions are constructed by
matching an interior wormhole geometry to an exterior
Schwarzschild vacuum. Analogously to their phantom and Chaplygin
counterparts, these VDW quintessence wormholes have far-reaching
astrophysical and cosmological consequences, such as the
production of closed timelike curves and the consequent violation
of causality.

This paper is outlined in the following manner: In Sec.
\ref{Sec:eq-structure}, the equations of structure for van der
Waals quintessence stellar models are presented. In Sec.
\ref{Sec:gravastar}, an exact solution of a gravastar is found,
and the respective properties and characteristics are analysed. In
Sec. \ref{Sec:wormhole}, an exact solution of a traversable
wormhole geometry is also found. In Sec. \ref{Conclusion}, we
conclude.

\section{VDW quintessence stars: Equations of
structure}\label{Sec:eq-structure}

Consider a static and spherically symmetric spacetime, given by
the following metric, in curvature coordinates
\begin{eqnarray}
ds^2=-e^{2\Phi(r)}\,dt^2+\frac{dr^2}{1- 2m(r)/r}
   +r^2 \,(d\theta ^2+\sin ^2{\theta} \, d\phi ^2) \label{metric}
\,,
\end{eqnarray}
where $\Phi(r)$ and $m(r)$ are arbitrary functions of the radial
coordinate, $r$. The function $m(r)$ is the quasi-local mass, and
is denoted as the mass function.

The Einstein field equation, $G_{\mu\nu}=8\pi T_{\mu\nu}$, where
$G_{\mu\nu}$ is the Einstein tensor and $T_{\mu\nu}$ the
stress-energy tensor, provides the following relationships
\begin{eqnarray}
m'&=&4\pi r^2 \rho   \label{mass}\,,\\
\Phi'&=&\frac{m+4\pi r^3 p_r}{r(r-2m)} \label{pressure}\,,\\
p_r'&=&-\frac{(\rho+p_r)(m+4\pi r^3 p_r)}{r(r-2m)}
+\frac{2}{r}(p_t-p_r)\label{anisotTOV}\,,
\end{eqnarray}
where the prime denotes a derivative with respect to the radial
coordinate, $r$. $\rho(r)$ is the energy density, $p_r(r)$ is the
radial pressure, and $p_t(r)$ is the lateral pressure measured in
the orthogonal direction to the radial direction. Equation
(\ref{anisotTOV}) corresponds to the anisotropic pressure
Tolman-Oppenheimer-Volkoff (TOV) equation, and may be obtained
using the conservation of the stress-energy tensor,
$T^{\mu\nu}_{\;\;\;\;;\nu}=0$.

The four-velocity of a static observer, at rest at constant $r,
\theta, \phi$, is $U^{\mu}=dx^{\mu}/{d\tau}
=(U^{\,t},0,0,0)=(e^{-\Phi(r)},0,0,0)$. The observer's
four-acceleration is $a^{\mu} =U^{\mu}{}_{;\nu}\,U^{\nu}$, so that
taking into account metric (\ref{metric}), we have $a^t=0$ and
\begin{eqnarray}
a^r=\Gamma^{r}{}_{tt}\,\left(\frac{dt}{d\tau}\right)^2
    ={\Phi}'\,(1-b/r)\,.          \label{radial-acc}
\end{eqnarray}
Note that from the geodesic equation, a radially moving test
particle, which initially starts at rest, obeys the following
equation of motion
\begin{equation}
\frac{d^2r}{d\tau^2}=-\Gamma^{r}{}_{tt}\left(\frac{dt}{d\tau}\right)^2=-a^r
\,.
      \label{radial-accel}
\end{equation}
$a^r$ is the radial component of proper acceleration that an
observer must maintain in order to remain at rest at constant
$r,\,\theta,\,\phi$. One may consider that the geometry is
attractive if $a^r>0$, i.e., observers must maintain an
outward-directed radial acceleration to keep from being pulled
into the star; and repulsive if $a^r<0$, i.e., observers must
maintain an inward-directed radial acceleration to avoid being
pushed away from the star. This distinction depends on the sign of
$\Phi'$, as is transparent from Eq.~(\ref{radial-acc}). In
particular, for a constant redshift function, $\Phi'(r)=0$, static
observers are also geodesic. Thus, the convention used is that
$\Phi'(r)$ is positive for an inwardly gravitational attraction,
and negative for an outward gravitational repulsion \cite{Roman}.

Despite of the fact that, in a cosmological context, the van der
Waals fluid is considered homogeneous, inhomogeneities may arise
through gravitational instabilities, resulting in a nucleation of
the cosmic fluid due to the respective density perturbations.
Thus, these solutions may possibly originate from density
fluctuations in the cosmological background. Consider the VDW
equation of state for an inhomogeneous spherically symmetric
spacetime, given by
\begin{equation}
p_r=\frac{\gamma \rho}{1-\beta \rho}-\alpha \rho^2  \,.
\label{vdWstar}
\end{equation}
In the present context of relativistic stellar models, we shall
consider the pressure in Eq. (\ref{vdWstar}) as the radial
pressure, $p_r$, and the tangential pressure may be obtained via
Eq. (\ref{anisotTOV}). Note that the limiting case $\alpha,\beta
\rightarrow 0$, reduces to the dark energy equation of state, with
$\gamma <-1/3$. Using Eqs. (\ref{mass})-(\ref{pressure}), then Eq.
(\ref{vdWstar}) provides the following relationship
\begin{equation}
r\left(1-\frac{2m}{r}\right) \Phi'=\frac{m}{r}+\frac{\gamma
m'}{1-\frac{\beta m'}{4 \pi r^2}} -\frac{\alpha m'^2}{4\pi r^2}
\,.
            \label{vdWEOS}
\end{equation}

Note that now we have four equations, Eqs.
(\ref{mass})-(\ref{anisotTOV}) and Eq. (\ref{vdWstar}), with five
unknown functions of $r$, i.e., $\Phi(r)$, $m(r)$, $\rho(r)$,
$p_r(r)$ and $p_t(r)$. To solve the system, one may adopt
different strategies, namely, one may model an appropriate
spacetime geometry by imposing $m(r)$ and/or $\Phi(r)$ by hand and
consequently determine the stress-energy tensor components. In
counterpart, one may consider an adequate source of the spacetime
geometry by imposing the stress-energy components, and
consequently determine the metric fields.

Another useful and interesting relationship may be obtained from
Eqs. (\ref{mass}), (\ref{anisotTOV}) and Eq. (\ref{vdWstar}),
which provide
\begin{eqnarray}
\Delta&=&\frac{1}{8\pi
r^2}\Bigg\{(m''r-2m')\left[\frac{\gamma}{\left(1-\frac{\beta
m'}{4\pi r^2}\right)^2}-\frac{\alpha m'}{2\pi r^2} \right]
    +m'r\left[1+\frac{\gamma}{1-\frac{\beta m'}{4\pi
r^2}}-\frac{\alpha m'}{4\pi r^2} \right]\Phi' \Bigg\}
   \label{TOVdark2}   \,.
\end{eqnarray}
$\Delta=p_t-p_r$ is denoted the anisotropy factor, as it is a
measure of the pressure anisotropy of the fluid comprising the VDW
quintessence star. The factor $\Delta/r$ represents a force due to
the anisotropic nature of the stellar model, which is repulsive,
i.e., being outward directed if $p_t>p_r$, and attractive if
$p_t<p_r$. $\Delta=0$ corresponds to the particular case of an
isotropic pressure VDW quintessence star.

One may, in principle, construct asymptotically flat spacetimes,
where $\Phi(r)\rightarrow 0$, and $m(r)/r \rightarrow 0$, as
$r\rightarrow \infty$. An alternative approach is to consider a
cut-off of the stress-energy tensor at a junction radius $a$. For
instance, consider for simplicity that the exterior solution is
the Schwarzschild spacetime, given by
\begin{eqnarray}
ds^2&=&-\left(1-\frac{2M}{r}\right)\,dt^2+
\left(1-\frac{2M}{r}\right)^{-1}dr^2
     +r^2(d\theta ^2+\sin ^2{\theta}\, d\phi ^2)
\label{metricvacuumlambda}  \,.
\end{eqnarray}
$M$ may be interpreted as the VDW quintessence star's total mass.
In this case the spacetimes given by the metrics Eq.
(\ref{metric}) and (\ref{metricvacuumlambda}) are matched at $a$,
and one has a thin shell surrounding the star. Using the
Darmois-Israel formalism, the surface stresses are given by
\begin{eqnarray}
 \sigma&=&-\frac{1}{4\pi a}
\left(\sqrt{1-\frac{2M}{a}+\dot{a}^2}-
\sqrt{1-\frac{2m(a)}{a}+\dot{a}^2} \, \right)
    \label{surfenergy}   ,\\
 {\cal P}&=&\frac{1}{8\pi a} \Bigg[\frac{1-\frac{M}{a}
+\dot{a}^2+a\ddot{a}}{\sqrt{1-\frac{2M}{a}+\dot{a}^2}}
     -\frac{(1+a\Phi') \left(1-\frac{2m}{a}+\dot{a}^2
\right)+a\ddot{a}-\frac{\dot{a}^2(m-m'a)}{(a-2m)}}{\sqrt{1-\frac{2m(a)}{a}+\dot{a}^2}}
\, \Bigg]         \,,
    \label{surfpressure}
\end{eqnarray}
where the overdot denotes a derivative with respect to the proper
time, $\tau$; $\sigma$ is the surface energy density and ${\cal
P}$ the surface pressure (see Refs. \cite{Lobo-WH} for details).
The static case is given by taking into account
$\dot{a}=\ddot{a}=0$. The total mass of the VDW quintessence star,
for the static case, is given by
\begin{equation}\label{totalmass}
M=m(a_0)+m_s(a_0)\left[\sqrt{1-\frac{2m(a_0)}{a_0}}-\frac{m_s(a_0)}{2a_0}\right]
\,,
\end{equation}
where $m_s$ is the surface mass of the thin shell, and is defined
as $m_s=4\pi a^2 \sigma$.

\section{VDW quintessence gravastars}\label{Sec:gravastar}

The Mazur-Mottola model is constituted by an onion-like structure
with five layers, including two thin-shells, with surface stresses
$\sigma_\pm$ and ${\cal P}_\pm$, where $\sigma$ is the surface
energy density and ${\cal P}$ is the surface tangential pressure.
The interior of the solution is replaced by a segment of de Sitter
space, which is then matched to a finite thickness shell of stiff
matter with the equation of state $p=\rho$. The latter is further
matched to an external Schwarzschild vacuum with $p=\rho=0$.
A simplification of the Mazur-Mottola configuration, and its
dynamic stability, was considered in Ref. \cite{VW} using a
three-layer solution \cite{VW}, i.e., a de Sitter interior
solution was matched to a Schwarzschild exterior solution at a
junction surface, comprising of a thin shell with surface stresses
$\sigma$ and ${\cal P}$.

Consider the specific case of a constant energy density,
$\rho(r)=\rho_0$, so that Eq. (\ref{mass}) provides the following
mass function
\begin{equation}
m(r)=A r^3 \,,
      \label{const-rho}
\end{equation}
where for simplicity, the definition $A=4\pi \rho_0/3$ is used.
Using Eq. (\ref{vdWEOS}), one finds the following expression
\begin{equation}
\Phi'(r)=\frac{Ar\chi}{1-2Ar^2}
 \,.
      \label{grav-Phi'}
\end{equation}
where, for notational convenience, the constant $\chi$ is defined
by
\begin{equation}
\chi=\left(1+\frac{3\gamma}{1-\frac{3\beta A}{4\pi}}-\frac{9\alpha
A}{4\pi}\right)  \,.
\end{equation}

Now, in order to obtain an accelerated behaviour of the Universe,
the condition $\rho+3p<0$, should be obeyed. We shall use this
condition, in the present inhomogeneous spacetime, which yields
the following inequality
\begin{equation}
\rho\left(1+\frac{3\gamma}{1-\beta \rho}-3\alpha\rho \right)<0 \,.
\label{acc-univ}
\end{equation}
Note that in order to provide a positive energy density, one may
also impose the following simultaneous conditions, in terms of the
VDW equation of state parameters
\begin{equation}
\rho>\frac{1}{\beta}, \qquad \frac{\beta+3\alpha}{\alpha\beta}>0,
\qquad \frac{(\beta+3\alpha)^2}{12\alpha\beta}>(1-3\gamma)
 \,.    \label{parameters}
\end{equation}
See Ref. \cite{Capob} for further details. Taking into account the
mass function given by Eq. (\ref{const-rho}), and using Eq.
(\ref{mass}), we have $\rho=\rho_0=3A/(4\pi)$, so that inequality
(\ref{acc-univ}) takes the following form
\begin{equation}
\frac{3A \chi}{4\pi}  <0
 \,.
\end{equation}
from which one readily verifies, using Eq. (\ref{grav-Phi'}), that
$\Phi'(r)<0$, providing the necessary repulsive character, which
is a fundamental property of gravastar models. Note that as we are
considering positive energy densities, so that $A>0$, in the
gravastar models, we have $\chi<0$.

Equation (\ref{grav-Phi'}), can be integrated to provide the
following spacetime metric
\begin{equation}
ds^2=-\left(1-2Ar^2\right)^{-\chi/2}\,dt^2+\frac{dr^2}{1-
2Ar^2}+r^2 \,(d\theta ^2+\sin ^2{\theta} \, d\phi ^2)
 \label{gravametric}\,.
\end{equation}

\begin{figure}[h]
\centering
  \includegraphics[width=2.8in]{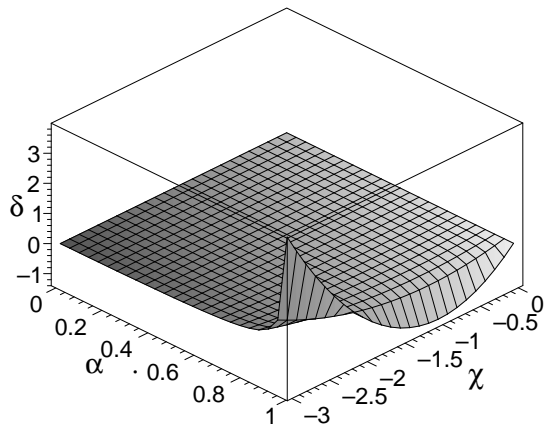}
  \caption{Plot of the anisotropy factor for a VDW quintessence star
  with a constant energy density. We have defined the following
  dimensionless parameters: $\delta=\Delta/A$ and $\alpha=\sqrt{A}r$.}
  \label{Fig:Delta}
\end{figure}

The stress-energy scenario is given by $\rho=\rho_0=3A/(4\pi)$ and
\begin{eqnarray}
p_r(r)&=&\frac{A(\chi-1)}{4\pi}  \,,  \\
p_t(r)&=&\frac{1}{8\pi}\left[\frac{A\chi(1+Ar^2\chi)}{1-2Ar^2}+A(\chi-2)\right]\,.
\end{eqnarray}
The anisotropy factor is provided by the following relationship
\begin{equation}
\Delta=\frac{A^2r^2\chi(2+\chi)}{8\pi(1-2Ar^2)} \,,
\end{equation}
which is plotted in Fig. \ref{Fig:Delta}. Note that $\Delta=0$,
i.e., $p_r=p_t$, at the center $r=0$ as was to be expected.

\section{VDW quintessence traversable wormholes}\label{Sec:wormhole}

The spacetime metric representing a spherically symmetric and
static wormhole is given by Eq. (\ref{metric}), where $\Phi(r)$
and $b(r)=2m(r)$ are arbitrary functions of the radial coordinate,
$r$, denoted as the redshift function, and the form function,
respectively \cite{Morris}. The radial coordinate has a range that
increases from a minimum value at $r_0$, corresponding to the
wormhole throat, to infinity.
To be a wormhole solution a flaring out condition of the throat is
imposed, i.e., $(b-b'r)/b^2>0$ \cite{Morris}. From this we verify
that at the throat $b(r_0)=r=r_0$, the condition $b'(r_0)<1$ is
imposed to have wormhole solutions. For the wormhole to be
traversable, one must demand that there are no horizons present,
which are identified as the surfaces with $e^{2\Phi}\rightarrow
0$, so that $\Phi(r)$ must be finite everywhere. Note that the
condition $1-b/r>0$ is also imposed.

Consider, for instance, the specific case of of $b(r)=r_0^2/r$,
which obeys the conditions of a wormhole, so that Eq.
(\ref{vdWEOS}) provides the following solution
\begin{eqnarray}
\Phi(r)&=&-\frac{1}{8\pi r_0^2
+\beta}\left[\frac{\alpha-\beta}{4}+\frac{\alpha \beta}{32\pi
r_0^2}-2\pi r_0^2(1-\gamma)\right]\,\ln(r^2-r_0^2)+\frac{\pi r_0^2
\gamma}{8\pi r_0^2 +\beta}\ln(8\pi r^4+\beta r_0^2)
      \nonumber    \\
 &&+\left(\frac{\alpha}{16\pi r_0^2}-\frac{1}{2}\right)\ln(r)
 -\frac{\alpha}{32\pi r^2}
       -\frac{\alpha r_0^2}{64\pi r^4}
     -\frac{\sqrt{2\pi \beta}\,r_0\gamma}{2(8\pi r_0^2 +\beta)}\;
\arctan\left(\frac{2r^2}{r_0}\sqrt{\frac{2\pi}{\beta}}\right)+C\,,
            \label{Phi}
\end{eqnarray}
where $C$ is a constant of integration. We verify the existence of
an event horizon, $r=r_0$, from the first term in the
right-hand-side. Thus, fine-tuning the $\gamma$ parameter as
\begin{equation}
\gamma=1-\frac{1}{(2\pi
r_0^2)}\left(\frac{\alpha-\beta}{4}+\frac{\alpha\beta}{32\pi
r_0^2}\right)  \,,
     \label{fine-tune}
\end{equation}
Eq. (\ref{Phi}) reduces to
\begin{eqnarray}
\Phi(r)&=&\left(\frac{\alpha}{16\pi
r_0^2}-\frac{1}{2}\right)\ln(r)+\frac{\pi r_0^2 \gamma}{8\pi r_0^2
+\beta}\ln(8\pi r^4+\beta r_0^2)
      \nonumber    \\
&&      -\frac{\sqrt{2\pi \beta}\,r_0\gamma}{2(8\pi r_0^2
+\beta)}\;
\arctan\left(\frac{2r^2}{r_0}\sqrt{\frac{2\pi}{\beta}}\right)
      -\frac{\alpha}{32\pi r^2}-\frac{\alpha r_0^2}{64\pi r^4}
      +C\,.
            \label{Phi2}
\end{eqnarray}
This now corresponds to a traversable  wormhole, as one verifies
the absence of an event horizon. Note that this solution is not
asymptotically flat, so that one needs to match the latter to an
exterior vacuum spacetime. If we consider a further imposition,
namely, $\alpha=8\pi r_0^2$, then Eq. (\ref{fine-tune}) becomes
zero, $\gamma=0$, so that the redshift function assumes a
particularly simple form, given by
\begin{equation}
 \Phi(r)=-\frac{r_0^2}{4r^2}\left(1+\frac{r_0^2}{2r^2}\right)+C
 \,.       \label{Phi3}
\end{equation}
The VDW equation of state, Eq. (\ref{vdWEOS}), for this particular
choice of parameters is $p_r=-8\pi r_0^2\, \rho^2$, implying a
negative radial pressure. Note that this specific choice of a
wormhole is asymptotically flat. The stress-energy components take
the following form
\begin{eqnarray}
\rho(r)&=&-\frac{r_0^2}{8\pi r^4}\,, \qquad
p_r(r)=-\frac{r_0^6}{8\pi r^8}
         \\
p_t(r)&=&-\frac{r_0^2\left(4r^8+r_0^2\, r^6-15r_0^4\, r^4+r_0^6\,
r^2+r_0^8 \right)}{32\pi r^{12}} \,.
\end{eqnarray}
This solution corresponds to a rather exotic form of a VDW fluid,
due to the presence of a negative energy density. The null energy
condition, defined as $T_{\mu\nu}\,k^\mu\,k^\nu>0$, where
$k^{\mu}$ any null vector, is also violated, i.e, $\rho+p_r<0$,
which is a necessary condition of wormholes.

It is also of interest to note that the VDW cosmic fluid with an
isotropic pressure may be completely determined by only one
parameter in the equation of state, so that one may consider
$\gamma$ as the only independent parameter needed to describe the
VDW fluid \cite{Capob}. Therefore, using a wide variety of values
of $\gamma$ constrained by observations considered in Ref.
\cite{Capob}, one may obtain restrictions on the parameters
$\alpha$ and $\beta$, through Eq. (\ref{fine-tune}), in the
inhomogeneous solutions considered here. On the other hand,
although the particular case of $\gamma=0$ is not included in the
best fits parameter range \cite{Capob}, it is not excluded from
the observational constraints. This should not be considered a
serious shortcoming, as the parameters $\alpha$, $\beta$ and
$\gamma$ are defined using the critical values of the energy
density, the isotropic pressure and the critical volume
\cite{Capob}, and in the presence of anisotropic pressures the
respective definitions should be extended to include the
transverse pressure. Despite of this fact, and due to the
analytical complexity of Eq. (\ref{Phi2}), we shall, for
simplicity, consider the case $\gamma=0$ and further analyze the
physical properties and characteristics of this specific wormhole
solution below, which prove to be extremely interesting.

One may also quantify the ``total amount'' of energy condition
violating matter, which amounts to calculating the ``volume
integral quantifier'' defined as $\int T_{\mu\nu}k^\mu k^\nu \,dV$
\cite{VKD1}. The amount of violation is defined as the extent to
which this integral becomes negative. Using the ``volume integral
quantifier'', which provides information about the ``total
amount'' of averaged null energy condition (ANEC) violating matter
in the spacetime (see Ref. \cite{VKD1} for details), given by
$I_V=\int\left[\rho(r)+p_r(r)\right]dV$, with a cut-off of the
stress-energy at $a$, we have
\begin{eqnarray}
I_V&=&\int_{r_0}^a
\left(r-b\right)\left[\ln\left(\frac{e^{2\Phi}}{1-b/r}\right)\right]'\;dr
\,.
    \label{vol-int}
\end{eqnarray}
Now, using the form function given by $b(r)=r_0^2/r$, and the
redshift function provided by Eq. (\ref{Phi3}), and evaluating the
integral, one finally ends up with the following simplified
expression for the ``volume integral quantifier''
\begin{eqnarray}
I_V=\frac{r_0(r_0^5+5r_0a^4-6a^5)}{5a^5} \,.
\end{eqnarray}
By taking the limit $a\rightarrow r_0$, one readily verifies that
$I_V\rightarrow 0$. It is also interesting to note that in the
limit $a\rightarrow \infty$, the volume integral tends to a finite
value, i.e., $I_V= -6r_0/5$. This proves that as in the specific
case of phantom \cite{phantomWH} and Chaplygin
\cite{Lobo-chaplygin} wormholes, one may theoretically construct a
wormhole with arbitrarily small amounts of a VDW quintessence
fluid. As emphasized in Ref. \cite{phantomWH}, this result is not
unexpected, however, it is interesting to note the relative ease
with which one may theoretically construct wormholes supported by
infinitesimal amounts of exotic fluids used in cosmology to
explain the present accelerated cosmic expansion.

It is also of a particular interest to analyze the traversability
conditions. We will be interested in specific solutions for
traversable wormholes and assume that a traveller of an absurdly
advanced civilization, with human traits, begins the trip in a
space station in the lower universe, at proper distance $l=-l_1$,
and ends up in the upper universe, at $l=l_2$. The proper distance
is given by $dl=\pm (1-b/r)^{-1/2}\,dr$. Assume that the traveller
has a radial velocity $v(r)$, as measured by a static observer
positioned at $r$. One may relate the proper distance travelled
$dl$, radius travelled $dr$, coordinate time lapse $dt$, and
proper time lapse as measured by the observer $d\tau$, by the
following relationships
\begin{eqnarray}
v&=&e^{-\Phi}\,\frac{dl}{dt}=\mp\;e^{-\Phi}\,
\left(1-\frac{b}{r}\right)^{-1/2}\frac{dr}{dt}  \,,
             \\
v\,\gamma&=&\frac{dl}{d\tau}=\mp\;
\left(1-\frac{b}{r}\right)^{-1/2}\frac{dr}{d\tau}  \,.
\end{eqnarray}
See \cite{Morris} for details.

For a convenient trip through the wormhole, certain conditions
should also be imposed~\cite{Morris}. Firstly, the entire journey
should be done in a relatively short time as measured both by the
traveller, $\Delta \tau_{\rm tr}$, and by observers who remain at
rest at the stations, $\Delta t_{\rm st}$. For simplicity, we
shall take into account non-relativistic and constant traversal
velocities, $\gamma \approx 1$. Thus, $\Delta \tau_{\rm tr}$ and
$\Delta t_{\rm st}$ are given by
\begin{eqnarray}
\Delta \tau_{\rm tr} &=&\int_{-l_1}^{+l_2}
\frac{dl}{v\gamma}=2\int_{r_0}^{a}
\frac{dr}{(1-b/r)^{1/2}\,v\gamma}
    \geq 2\int_{r_0}^{a} \frac{dr}{v\gamma} \approx
\frac{2(a-r_0)}{v}
\label{travelertime}, \\
\Delta t_{\rm st} &=&\int_{-l_1}^{+l_2} \frac{dl}{v
e^{\Phi}}=2\int_{r_0}^{a} \frac{dr}{(1-b/r)^{1/2}\,v\,e^{\Phi}}
     \geq 2\int_{r_0}^{a} \frac{dr}{v e^{\Phi}} =
\frac{2}{v}\int_{r_0}^{a}  e^{-\Phi}\,dr
 \label{observertime},
\end{eqnarray}
Considering the redshift function provided by Eq. (\ref{Phi3}),
and $a=2r_0$, so that $\int_{r_0}^{a} e^{-\Phi}\,dr \approx 1.18
\,r_0$, we verify that both $\Delta \tau_{\rm tr}$ and $\Delta
t_{\rm st}$ are bounded from below by $\sim 2r_0/v$.

An important traversability condition required is that the
acceleration felt by the traveller should not exceed Earth's
gravity \cite{Morris}. Thus, the traveller's four-acceleration
expressed in his proper reference frame, considering once again
non-relativistic and constant traversal velocities, $\gamma
\approx 1$, yields the following restriction (see Ref.
\cite{Morris} for details)
\begin{equation}
|\vec{a}|\approx \left |\left (1-\frac{b}{r}\right)^{1/2}
\Phi'\right| \leq g_{\oplus} \label{acceleration} \,,
\end{equation}
which is readily verified at the throat, where the conditions are
most severe.

Another important condition is that an observer traversing through
the wormhole should not be ripped apart by enormous tidal forces.
Thus, it is required that the tidal accelerations felt by the
traveller should not exceed, for instance, the Earth's
gravitational acceleration \cite{Morris}. The constraint $|\Delta
a^{\hat{\mu}'}|\leq g_{\oplus}$ provides the tidal acceleration
restrictions as measured by a traveller moving radially through
the wormhole, given by the following inequalities
\begin{eqnarray}
\left |\left (1-\frac{b}{r} \right ) \left [\Phi ''+(\Phi ')^2-
\frac{b'r-b}{2r(r-b)}\Phi' \right] \right
|\,\big|\eta^{\hat{1}'}\big| \leq  g_\oplus   \,,
    \label{radialtidalconstraint}    \\
\left | \frac{\gamma ^2}{2r^2} \left [v^2\left (b'-\frac{b}{r}
\right )+2(r-b)\Phi ' \right] \right | \,\big|\eta^{\hat{2}'}\big|
\leq   g_\oplus    \,.    \label{lateraltidalconstraint}
\end{eqnarray}
We refer the reader to Ref. \cite{Morris} for details related to
the deduction of these conditions. The factors
$\big|\eta^{\hat{1}'}\big|$ and $\big|\eta^{\hat{2}'}\big|$
appearing in inequalities
(\ref{radialtidalconstraint})-(\ref{lateraltidalconstraint}) are
the spatial separations between radial and lateral parts of the
traveller's body, as measured in his proper reference frame. For
computational purposes we may assume
$\big|\eta^{\hat{1}'}\big|\approx \big|\eta^{\hat{2}'}\big|\approx
2\,{\rm m}$.

The radial tidal constraint, Eq. (\ref{radialtidalconstraint}),
constrains the redshift function, and the lateral tidal
constraint, Eq. (\ref{lateraltidalconstraint}), constrains the
velocity with which observers traverse the wormhole. These
inequalities are particularly simple at the throat, $r_0$, and
reduce to
\begin{eqnarray}
|\Phi '(r_0)| \leq
\frac{2g_{\oplus}\,r_0}{(1-b')\,|\eta^{\hat{1}'}|} \,, \label{1st}
          \\
\gamma^2 v^2 \leq
\frac{2g_{\oplus}\,r_0^2}{(1-b')\,|\eta^{\hat{2}'}|}    \,.
\label{2nd}
\end{eqnarray}
Taking into account the redshift function given by Eq.
(\ref{Phi3}), and the form function considered above, i.e.,
$b(r)=r_0^2/r$, we verify that evaluated at the throat we have
$\Phi'(r_0)=1/r_0$ and $b'(r_0)=-1$. Thus, inequality (\ref{1st})
provides the restriction $1 \leq g_{\oplus}
r_0^2/|\eta^{\hat{1}'}|$. Considering the minimum value, i.e.,
$g_{\oplus} r_0^2/|\eta^{\hat{1}'}|\approx 1$, and that
$|\eta^{\hat{1}'}|\approx |\eta^{\hat{2}'}|\approx 2{\rm m}$, we
verify that the minimum value of the throat is given by
$r_0\approx 2\times 10^{8}\,{\rm m}$. Using the latter
approximation, and taking into account non-relativistic and
constant traversal velocities, $\gamma \approx 1$, then inequality
(\ref{2nd}) provides the following restriction for the velocity,
$v \leq c$. For instance, considering that the traversal velocity
be of the order $v\sim 10^{-2}\,c$, and assuming that the
traversal times take their minimum value, we have $\Delta
\tau_{\rm tr}\sim \Delta t_{\rm st}\sim 10^2\,{\rm s}$. This
provides an extremely reasonable traversable wormhole, useful for
travellers with human traits.

\section{Conclusion}\label{Conclusion}

In conclusion, noting that the VDW quintessence equation of state
is an interesting scenario for describing the late universe, we
have explored the construction of relativistic stellar models, in
particular gravastars and traversable wormholes, supported by a
VDW equation of state. These solutions may possibly originate from
density fluctuations in the cosmological background, resulting in
the nucleation through the respective density perturbations. We
have found an exact solution of a gravastar model, by considering
the specific case of a constant energy density. Relatively to
traversable wormhole geometries, we also found an exact solution
of an asymptotically flat wormhole, as well as solutions, where
the exotic matter is constrained to the throat neighborhood, by
considering a matching of an interior wormhole geometry to an
exterior Schwarzschild vacuum solution. It was also found that one
may theoretically construct a wormhole with arbitrarily small
amounts of a VDW quintessence fluid. Several characteristics and
properties of these traversable wormholes using the traversability
conditions were also explored.

\section*{Acknowledgements}
This work was partially funded by Funda\c{c}\~{a}o para a
Ci\^{e}ncia e Tecnologia (FCT) -- Portugal through the grant
SFRH/BPD/26269/2006.



\end{document}